\documentclass[twocolumn,amsmath,amssymb,osajnl]{revtex4}
\usepackage{epsfig}
\usepackage{mathrsfs}

\newcommand{\rrr}{\mathbf{r}}
\newcommand{\EEE}{\mathbf{E^{{}}}}
\newcommand{\DDD}{\mathbf{D^{{}}}}
\newcommand{\nablabf}{\boldsymbol{\nabla}}

\begin{document}

\title{Semi-analytical approach to short-wavelength
dispersion and modal properties of photonic crystal fibers}

\author{Niels Asger Mortensen}

\affiliation{MIC -- Department of Micro and Nanotechnology,\\
Technical University of Denmark, DK-2800 Kongens Lyngby, Denmark }

\begin{abstract}
We consider photonic crystal fibers made from arbitrary base
materials and derive a unified semi-analytical approach for the
dispersion and modal properties which applies to the
short-wavelength regime. In particular we calculate the dispersion
and the effective index and comparing to fully-vectorial plane
wave simulations we find excellent agreement. We also calculate
asymptotic results for the mode-field diameter and the
$V$-parameter and from the latter we predict that the fibers are
endlessly single mode for a normalized air-hole diameter smaller
than $0.42$, independently of the base material.
\end{abstract}


\maketitle

Photonic crystal fibers (PCF) are dielectric optical fibers with
an array of air holes running along the full length of the fiber.
Typically, the fibers employ a single dielectric base material
(with dielectric function $\varepsilon_b=n_b^2$) and historically
silica has been the most common choice.\cite{knight1996,birks1997}
Recently other base materials have been studied including
chalcogenide glass, lead silicate glass, telluride glass, bismuth
glass, silver halide, teflon, and plastics/polymers. The
fabricated fibers typically share the same overall geometry with
the air holes arranged in a triangular lattice and the core defect
being formed by the removal of a single air-hole. There has been a
major theoretical and numerical effort to understand the
dispersion and modal properties of especially silica-based PCFs,
but the scale-invariance of Maxwell's equations\cite{joannopoulos}
cannot be applied directly to generalize the results to other base
materials. The reason is that PCFs made from different base
materials do not relate to each other by a linear scaling of the
dielectric function, $\varepsilon(\rrr)\nrightarrow
s^2\varepsilon(\rrr)$, with $s$ being a positive scalar. The
increased focus on use of new base materials thus calls for a
theory of PCFs with an arbitrary base material.

This Letter offers a unified approach to the dispersion properties
which utilizes that the base material typically has a dielectric
function exceeding that of air significantly, $\varepsilon_b\gg
1$. The short-wavelength regime is characterized by having the
majority of the electrical field residing in the high-index base
material while the fraction of electrical field in the air holes
is vanishing, see e.g. Refs.~\onlinecite{riishede2003a}. The
calculation starts from the fully-vectorial wave-equation for the
electrical field,\cite{joannopoulos}
\begin{equation}\label{eq:EEE}
\nablabf\times\nablabf\times\EEE(\rrr) =
\varepsilon(\rrr)\frac{\omega^2}{c^2}\:\EEE(\rrr).
\end{equation}
For a fiber geometry with $z$ along the fiber axis we have
$\varepsilon(\rrr)=\varepsilon(x,y)$ and we look for solutions of
the plane-wave form $e^{i(\beta z-\omega t)}$ with the goal of
calculating the dispersion relation $\omega(\beta)$. The above
discussion for $\varepsilon_b\gg 1$ suggests that we can
approximate the problem by imposing the boundary condition that
$\EEE$ is zero at the interfaces to the air holes. Since the
displacement field $\DDD=\varepsilon\EEE$ is divergence free we
have $0=\varepsilon \nablabf\cdot\EEE+ \EEE\cdot\nablabf
\varepsilon \approx \varepsilon \nablabf\cdot\EEE$. In the bulk
matrix material the latter equality is exact and the wave equation
reduces to
\begin{equation}\label{eq:EEEapprox}
-\nablabf^2\EEE(\rrr) =
\varepsilon_b\frac{\omega^2}{c^2}\:\EEE(\rrr),
\end{equation}
which we solve with the boundary condition that $\EEE$ is zero at
the interfaces to the air holes. For the boundary condition to be
meaningful it is crucial that $\varepsilon_b\gg 1$ so that the
fraction of electrical field in the air holes is vanishing in the
short-wavelength regime. The wave problem has now become scalar,
but it should be emphasized that the approach is very different
from the usual scalar treatment\cite{snyder} and recent
applications to PCFs in the short-wavelength regime
\cite{birks1997,riishede2003a,birks2004} which take the electrical
field in the air holes into account. Obviously, the scalar problem
posed by Eq.~(\ref{eq:EEEapprox}) is separable and formally we
have that
\begin{equation}\label{eq:omega(beta)}
\omega=\sqrt{\Omega_{xy}^2+\Omega_z^2}=\frac{c}{n_b}\sqrt{\gamma^2\Lambda^{-2}+\beta^2}
\end{equation}
where $\Omega_z=c\beta/n_b$ is the frequency associated with the
longitudinal plane-wave propagation, $\Omega_{xy}=\gamma\times c
\Lambda^{-1}/n_b$ is the frequency associated with the transverse
confinement/localization, and $\gamma$ is a corresponding
dimensionless and purely geometrical number, which only depends on
the normalized air-hole diameter $d/\Lambda$. From
Eq.~(\ref{eq:EEEapprox}) it follows that $\gamma$ is an eigenvalue
governed by a scalar two-dimensional Schr\"{o}dinger-like equation
\begin{equation}\label{eq:schroding}
-\Lambda^2(\partial_x^2+\partial_y^2)\psi(x,y)=\gamma^2\psi(x,y),
\end{equation}
with the scalar function $\psi$ being subject to hard-wall
boundary conditions at the interfaces to the air-holes, {\it i.e.}
$\psi=0$ in the air holes.

\begin{figure}[t!]
\begin{center}
\epsfig{file=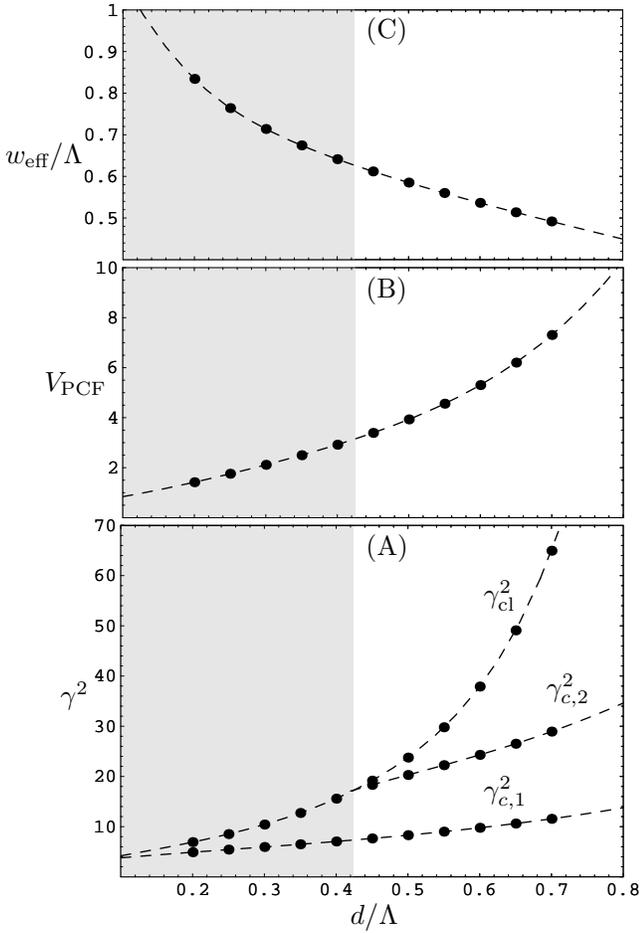, width=\columnwidth,clip,angle=0}
\end{center}
\caption{Panel (A) shows the geometrical eigenvalues $\gamma^2$
for the fundamental core ($c,1$), the second order core ($c,2$)
and the fundamental cladding (cl) modes versus normalized air-hole
diameter $d/\Lambda$, Panel (B) shows the corresponding
$V$-parameter, and Panel (C) the effective mode-field radius of
the fundamental core mode. The data-points are obtained from
finite-element simulations\cite{Comsol} of
Eq.~(\ref{eq:schroding}) and the dashed lines are guides to the
eyes. The gray region indicates the endlessly single-mode regime
with $V_\textrm{PCF}<\pi$.} \label{fig1}
\end{figure}

\begin{figure}[b!]
\begin{center}
\epsfig{file=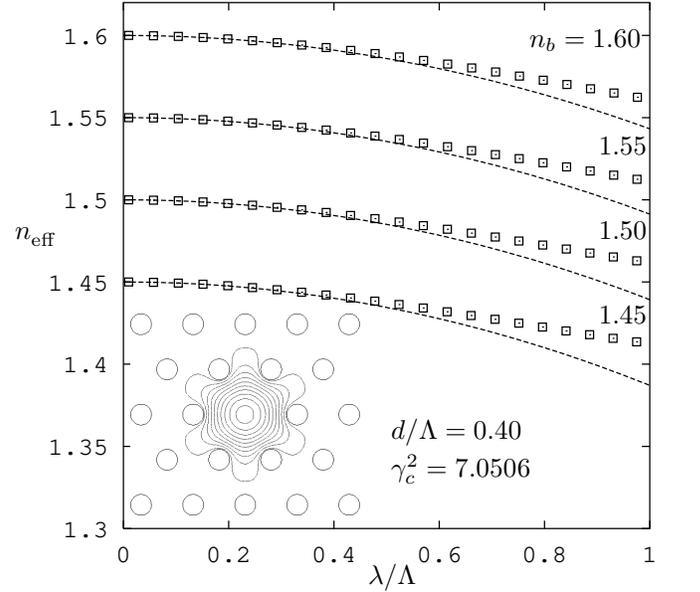, width=\columnwidth,clip,angle=0}
\end{center}
\caption{Effective index $n_\textrm{eff}$ of the fundamental core
mode versus normalized wavelength $\lambda/\Lambda$ for holey
fibers with normalized air-hole diameter $d/\Lambda=0.4$ and
varying base material. The dashed lines are the predictions of
Eq.~(\ref{eq:neff}) and the data points are results of
fully-vectorial plane-wave simulations.\cite{johnson2001} The
inset shows the fiber geometry and the fundamental core
eigenfunction $\psi_c$ with $\gamma_c^2=7.0506$, obtained with the
aid of a finite-element simulation.\cite{Comsol} } \label{fig2}
\end{figure}

\begin{figure}[t!]
\begin{center}
\epsfig{file=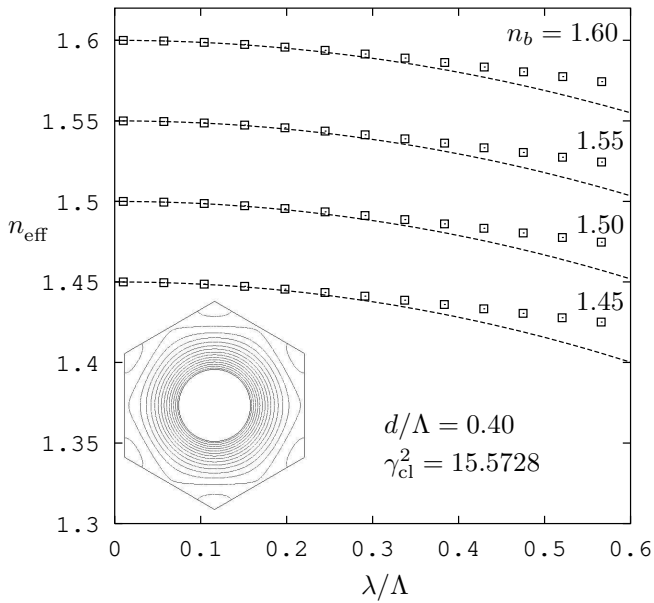, width=\columnwidth,clip,angle=0}
\end{center}
\caption{Effective index $n_\textrm{eff}$ of the fundamental
cladding mode versus normalized wavelength $\lambda/\Lambda$ for
holey fibers with normalized air-hole diameter $d/\Lambda=0.4$ and
varying base material. The dashed lines are the predictions of
Eq.~(\ref{eq:neff}) and the data points are results of
fully-vectorial plane-wave simulations.\cite{johnson2001} The
inset shows the unit cell of the periodic cladding structure and
the fundamental cladding eigenfunction $\psi_\textrm{cl}$ with
$\gamma_\textrm{cl}^2=15.5728$, obtained with the aid of a
finite-element simulation.\cite{Comsol} } \label{fig3}
\end{figure}

The developments in computational physics and engineering have
turned numerical solutions of partial differential equations in
the direction of a standard task. Here, we employ a finite-element
approach\cite{Comsol} to numerically solve
Eq.~(\ref{eq:schroding}) and calculate $\gamma^2$ versus
$d/\Lambda$. Panel (A) in Fig.~\ref{fig1} summarizes the results
for the fundamental core mode (see inset in Fig.~\ref{fig2}), the
second-order core mode, and the fundamental cladding mode (see
inset in Fig.~\ref{fig3}). For the core modes the problem has been
truncated by considering a sufficiently large domain with
Dirichlet boundary conditions while for the cladding modes we have
considered the unit-cell with periodic boundary conditions which
for symmetry reasons can be formulated in terms of Neumann
conditions.\cite{birks1997} As the normalized air-hole diameter
$d/\Lambda$ is increased the localization becomes more tight and
as expected the eigenvalue increases. For the fundamental core
mode the dashed line shows a third-order polynomial,
\begin{equation}\label{eq:polyfit}
\gamma_c^2\simeq {\mathscr A} + {\mathscr B}\:\frac{d}{\Lambda} +
{\mathscr C}\:\left(\frac{d}{\Lambda}\right)^2 + {\mathscr
D}\:\left(\frac{d}{\Lambda}\right)^3,
\end{equation}
with ${\mathscr A}=2.67$, ${\mathscr B}=12.51$, ${\mathscr
C}=-9.45$, and ${\mathscr D}=13.88$ being fitting parameters.

The recently proposed $V$-parameter\cite{mortensen2003c}
$V_\textrm{PCF}\equiv\Lambda
(\beta_c^2-\beta_\textrm{cl}^2)^{1/2}$ becomes
\begin{equation}
\lim_{\lambda\ll\Lambda}
V_\textrm{PCF}=\sqrt{\gamma_\textrm{cl}^2-\gamma_c^2}
\end{equation}
and the numerical results shown in Panel (B) agrees nicely with
the short-wavelength asymptotic limit of recent simulations on
silica-based PCFs.\cite{nielsen2003c} The endlessly single-mode
regime,\cite{birks1997} defined by
$V_\textrm{PCF}<\pi$,\cite{mortensen2003c} exists for
$d/\Lambda\lesssim 0.42$ independently of the base material. For
more detail on the modal cut-off see
Ref.~\onlinecite{mortensen2003c} and references therein. Panel (C)
shows results for the effective mode-field
radius\cite{nielsen2003b} $w_\textrm{eff}$ calculated from
$A_\textrm{eff}\equiv \pi w_\textrm{eff}^2$ with the effective
area given by
\begin{equation}
\lim_{\lambda\ll\Lambda}A_\textrm{eff}=\frac{\int dx dy\:
\big|\psi(x,y)\big|^2\int dx' dy'\: \big|\psi(x',y')\big|^2}{\int
dx dy\: \big|\psi(x,y)\big|^4}.
\end{equation}
As expected the mode-field diameter decreases as the normalized
air-hole diameter is increased and the mode becomes more
localized. For $V_\textrm{PCF}=\pi$ we find that
$w_\textrm{eff}/\Lambda\simeq 0.627$.

With Eqs.~(\ref{eq:omega(beta)}) and (\ref{eq:polyfit}) at hand we
have now provided a unified theory of the dispersion relation in
the short-wavelength regime for PCFs with arbitrary base materials
and Eq.~(\ref{eq:omega(beta)}) illustrates how geometrical
confinement modifies the linear free-space dispersion relation.

In fiber optics it is common to express the dispersion properties
in terms of the effective index $n_\textrm{eff}=c\beta/\omega$
versus the free-space wavelength $\lambda=c2\pi/\omega$. From
Eq.~(\ref{eq:omega(beta)}) it follows straightforwardly that
\begin{equation}\label{eq:neff}
n_\textrm{eff}=n_b\sqrt{1-\frac{\gamma^2}{4\pi^2n_b^2}\left(\frac{\lambda}{\Lambda}\right)^2}
\end{equation}
which obviously is in qualitative agreement with the accepted view
that $n_\textrm{eff}$ increases monotonously with decreasing
wavelength and approaches $n_b$ in the asymptotic short-wavelength
limit as reported for {\it e.g.} silica-based
PCFs.\cite{birks1997} However, how good is the quantitative
agreement for different base materials? In Figs.~\ref{fig2} and
\ref{fig3} we employ fully-vectorial plane-wave
simulations\cite{johnson2001} to compare Eq.~(\ref{eq:EEE}) with
the predictions of Eq.~(\ref{eq:neff}) for the fundamental core
and cladding modes, respectively. For the core-modes we have
employed a sufficiently large super-cell configuration. As seen
there is an over-all good agreement between the fully-vectorial
numerical results from Eq.~(\ref{eq:EEE}) and the semi-analytical
predictions of Eq.~(\ref{eq:neff}). In the short-wavelength limit
$\lambda\ll \Lambda$ the agreement is excellent, which underlines
the high relevance of the present results to large-mode area PCFs.
As the index of the base material is increased the agreement is
even better (results not shown). As the wavelength $\lambda$ is
increased and becomes comparable to the pitch $\Lambda$ the
quantitative agreement is less good. The reason is well-known from
small-core silica-based PCFs where a non-negligible fraction of
the electrical field is forced to reside in the air-hole regions
and where also vectorial effects of Eq.~(\ref{eq:EEE}) start to
matter.

In conclusion we have shown how a unified description of the
short-wavelength dispersion and modal properties is possible. The
theory illustrates how the waveguide dispersion originates from
the geometrical transverse localization of the mode and the
semi-analytical description of the short-wavelength properties is
readily applied to PCFs made from any base material.

\vspace{5mm} This work is financially supported by The Danish
Technical Research Council (Grant No.~26-03-0073). N.~A.
Mortensen's e-mail address is asger@mailaps.org.


\end{document}